\documentclass{elsart4}

\usepackage{epsfig}

\usepackage{amssymb}

\begin{document}

\begin{frontmatter}

\vspace*{-0.7cm}

\title{Magnetic excitations in the stripe phase
of high-$T_c$ superconductors}

\author[aff1]{G.S. Uhrig\corauthref{cor1}}
\ead{gu@thp.uni-koeln.de}
\corauth[cor1]{Tel:  +49/221/4703481; fax: +49/221/4705159}
\author[aff1]{K.P. Schmidt}
\author[aff2]{M. Gr\"{u}ninger}
\address[aff1]{Institut f\"{u}r Theoretische Physik, Universit\"{a}t zu K\"{o}ln,
Z\"{u}lpicher Str.\ 77, 50937 K\"{o}ln, Germany}
\address[aff2]{II. Physikalisches Institut, Universit\"{a}t zu K\"{o}ln,
Z\"{u}lpicher Str.\ 77, 50937 K\"{o}ln, Germany}




\begin{abstract}
The magnetic excitations in the stripe phase of high-$T_c$
superconductors are investigated in a model of spin ladders which are
effectively coupled via charged stripes.
Starting from the effective single-triplon model for the isolated
spin ladder, the quasi-one-dimensional
spin system can be described straightforwardly.
Very good agreement is obtained with recent neutron scattering data on
La$_{15/8}$Ba$_{1/8}$CuO$_4$ (no spin gap) and YBa$_2$Cu$_3$O$_{6.6}$
(gapped).
The signature of quasi-one-dimensional spin physics in a single-domain
stripe phase is predicted.
\end{abstract}

\end{frontmatter}


The role of magnetic excitations in the mechanism of high-$T_c$
superconductivity is still an unsettled issue, for a review see
e.g.\ Ref.\ \cite{norma03}.
Experimentally, a direct probe of these magnetic excitations
is inelastic neutron scattering
(INS) which provides information resolved both in energy and in momentum,
see e.g.\ Ref.\ \cite{sidis04}.
Three main features have to be understood:
(i) the so-called resonance peak \cite{norma03,sidis04}, which appears in the
superconducting phase at the antiferromagnetic wave vector
${\bf Q}_{\rm AF}=(1/2,1/2)$;
(ii) the appearance of superstructure satellites, which
are usually attributed to stripes \cite{norma03,schul89,zaane89,tranq95};
(iii) incommensurate excitations which lie energetically
both below and above the resonance mode \cite{bourg00,rezni04,mook02,arai99}.
There is growing evidence that these phenomena \cite{mook02,arai99,mook98}
are linked. Two recent papers  show the momentum dependence of the magnetic
excitations in the stripe-ordered \cite{hayde04} and in the superconducting
\cite{tranq04a} phase over a broad energy range. The data show stunning
similarities and allow a quantitative comparison with theory.

In the case of static stripes, the hole-poor regions are described
by spin ladders \cite{tranq95}. But the number of legs of these
spin ladders is not yet unambiguously determined.
Recent {\it ab initio} results suggest
that two-leg spin ladders are particularly stable \cite{anisi04}.
This is appealing since two-leg ladders are very well understood
\cite{barne93,trebs00,knett01b}. We have shown recently that
a model of weakly coupled spin ladders allows to describe the momentum
and the energy dependence of the neutron scattering intensity
of stripe-ordered La$_{15/8}$Ba$_{1/8}$CuO$_4$ (LBCO) \cite{tranq04a}
\emph{quantitatively} \cite{uhrig04a}. Qualitatively similar results
are obtained by considering coupled dimers \cite{vojta04} or by starting
from a N\'eel state \cite{kruge03,carls04}.
Approaches breaking the spin symmetry imply optical branches
\cite{kruge03,carls04,seibo04}.

We consider a spin-only model of undoped two-leg $S=1/2$ ladders separated
from each other by hole-rich bond-centered stripes (cf.\ Fig.\ 1a in Ref.\
\cite{uhrig04a}). Such a spin-only model certainly provides a useful
description if the charge excitations are gapped.
In the metallic stripe phase, the charge degrees of freedom will cause a
certain damping of the magnetic excitations, but recent results
indicate that this damping does not change the main physics \cite{seibo04}.
The Hamiltonian for a single ladder reads
$$
H= \sum_i [J_\perp {\bf S}_{i}^{\rm L}\cdot{\bf S}_{i}^{\rm R}
+J_\parallel({\bf S}_{i}^{\rm L}\cdot{\bf S}_{i+1}^{\rm L}+
{\bf S}_{i}^{\rm R}\cdot{\bf S}_{i+1}^{\rm R}
)] +H_{\rm cyc}
$$
where $i$ labels the rungs and $R$, $L$ the legs.
We use $J=J_\parallel \! = \! J_\perp$
since the system is derived from a square lattice.
Inclusion of the cyclic exchange
\begin{eqnarray*}
H_{\rm cyc} &=& J_{\rm cyc}
\sum_i[({\bf S}_{i}^{\rm L}\cdot{\bf S}_{i}^{\rm R})
({\bf S}_{i+1}^{\rm L}\cdot{\bf S}_{i+1}^{\rm R}) + \\
&& \hspace{-4mm}
({\bf S}_{i}^{\rm L}\cdot{\bf S}_{i+1}^{\rm L})
({\bf S}_{i}^{\rm R}\cdot{\bf S}_{i+1}^{\rm R}) -
 ({\bf S}_{i}^{\rm L}\cdot{\bf S}_{i+1}^{\rm R})
({\bf S}_{i+1}^{\rm L}\cdot{\bf S}_{i}^{\rm R})
]
\end{eqnarray*}
is justified both from first principles, e.g.\ \cite{mulle02a},
and phenomenologically, e.g.\ \cite{nunne02}. The established size
is $x_{\rm cyc} \! = \! J_{\rm cyc}/J \! = \! 0.20-0.25$, which is important
for quantitative agreement \cite{uhrig04a}. The ladders are coupled
\emph{ferromagnetically} between one another by $J'<0$
because the effective superexchange via a strongly doped stripe prefers
parallel alignment. This ferromagnetic coupling
shifts  the minima of the dispersion away from ${\bf Q}_{\rm AF}$
thus leading to incommensurate satellites \cite{uhrig04a}.

\begin{figure}[tb]
\centerline{\psfig{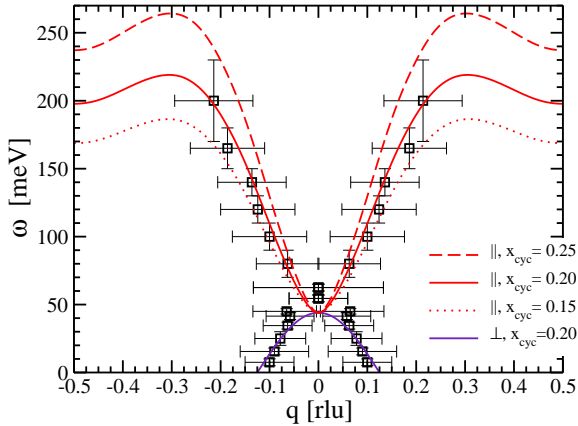}}
\caption{Triplon dispersion;
$q$ is the distance from ${\bf Q}_{\rm AF}$.
Symbols with error bars are INS data for LBCO
\cite{tranq04a}. Theoretical curves for
$(x_{\rm cyc},J{\rm [meV]},J'/J)=
(0.15,102,-0.098)$; $(0.20,127,-0.072)$ and $(0.25,162,-0.051)$
\cite{uhrig04a}.
Dispersion parallel (perpendicular) to the ladders is denoted $||$ ($\perp$).
Below the resonance mode curves for different $x_{\rm cyc}$ are
indistinguishable.}
\label{dispersion}
\end{figure}

The effective model for isolated ladders has been obtained
previously \cite{knett01b} by a continuous unitary transformation.
The elementary $S=1$ excitations are called triplons \cite{schmi03c}.
The ladders are coupled among themselves by $J'$ via a Bogoliubov
transformation.
It is only at this last step that the hard-core repulsion is neglected
\cite{uhrig04a}.

In Ref.\ \cite{uhrig04a} we compared the momentum dependence in constant
energy slices and the frequency dependence of the
momentum-integrated structure factor $S(\omega)$ with the INS data of
stripe-ordered LBCO \cite{tranq04a}. In Fig.\ \ref{dispersion} we show that
the same parameters determined before \cite{uhrig04a}
to describe $S(\omega)$ yield also an excellent description of the
dispersion. In particular the data for $x_{\rm cyc}\!=\!0.2$ agree very
well over the full energy range.
This strongly supports coupled ladders as model for
the magnetic excitations in the stripe phase.
The relatively large experimental error bars,
however, still leave room for possible
other features like a local `roton'-minimum \cite{seibo04}.
Certainly, further progress in experiment will settle this
question

\begin{figure}[t]
\centerline{\psfig{figure=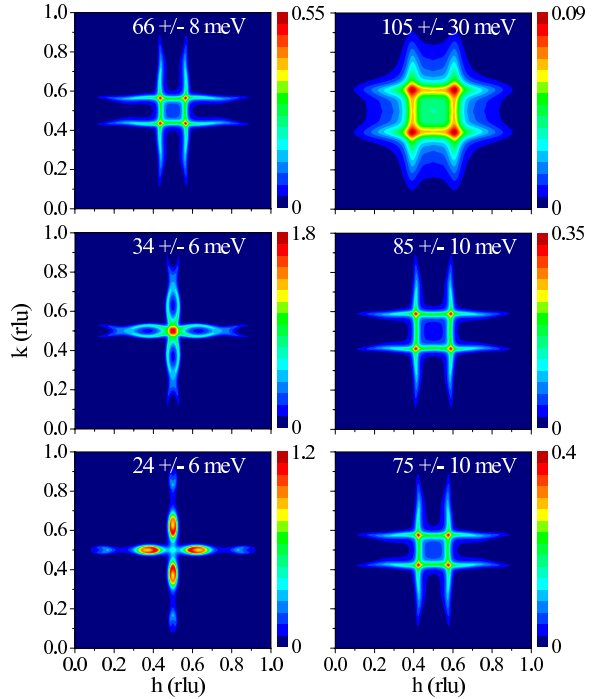,width=7.6cm}}
\caption{Constant-energy slices
for the indicated energies and resolutions from coupled ladders
(superposition of vertical and horizontal stripes)
for $x_{\rm cyc}\!=\!0.2, J\!=\!114{\rm meV}, J'\!=\!-0.035J$
\cite{uhrig04a};
to be compared with INS data of YBCO$_{6.6}$ \cite{hayde04}.
}
\label{ybco}
\end{figure}

An essential issue is how far
the magnetic excitations of the stripe
phase and of the superconducting state are related to each other.
The main features observed in the superconducting phase -- the resonance
mode, the downward dispersion below the resonance and the upward dispersion
above \cite{bourg00,rezni04,mook02,arai99,hayde04} -- are generic features
of our model \cite{uhrig04a}.
In underdoped YBa$_2$Cu$_3$O$_{7-\delta}$ (YBCO$_{7-\delta}$),
the observation of incommensurate scattering below the resonance has been
interpreted as a signature of stripe formation \cite{mook02,arai99,mook98}.
The new experimental results for underdoped YBCO$_{6.6}$ \cite{hayde04} are
stunningly similar to the data of stripe-ordered LBCO \cite{tranq04a} over
a broad energy range.

Fig.\ \ref{ybco} displays the results of our model for parameters pertaining
to YBCO$_{6.6}$ \cite{hayde04}. We stick to $x_{\rm cyc}\!=\!0.2$ and
determine $J\!=\!114$\,meV and $J'/J\!=\!-0.035$ via the experimental values
for the energy of the saddle point (i.e.\ the resonance)
$\omega_r\!=\!34$\,meV and the spin gap $\Delta \! = \! 20$\,meV \cite{dai99}.
We neglect the bilayer structure of YBCO, since the small bilayer coupling of
$\approx \! 0.1J$ \cite{rezni96} will not change the result for the
acoustic (odd) modes qualitatively.

On the one hand, neglecting the charge degrees of freedom appears to be a
much more severe shortcoming in superconducting YBCO than in stripe-ordered
LBCO. On the other hand, the use of the Bogoliubov transformation for
the interladder hopping and the concomitant omission of the hard-core
constraint is even better justified in YBCO
because of the finite spin gap and the smaller value of $J'/J \! =\! -0.035$.
It is not astounding that $J'$ varies from system to system. It is an
effective parameter which depends on the properties of the charges like
the doping level, the size of the charge gap and
the nature of the charge order.

Fig.\ \ref{ybco} agrees surprisingly well with the INS data of YBCO$_{6.6}$
\cite{hayde04}. The resonance mode at $\omega_r$ and the positions of the four
incommensurate peaks below and above $\omega_r$ are reproduced very well.
The general agreement strongly supports the underlying assumption that the
magnetic excitations can be described by coupled two-leg spin ladders.

\begin{figure}[t]
\centerline{\psfig{figure=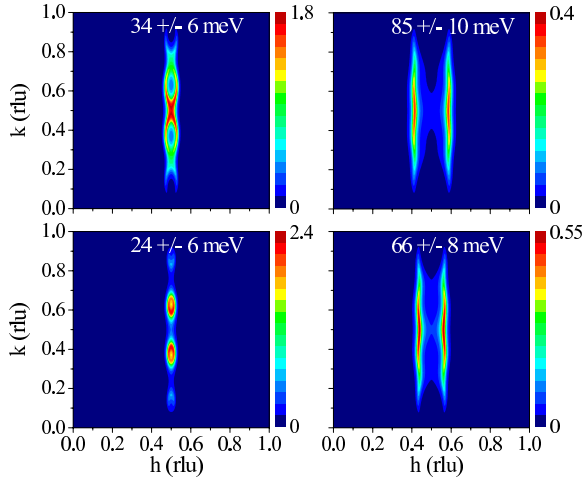,width=7.6cm}}
\caption{
Like Fig.\ \ref{ybco} for a single-domain stripe phase.
}
\label{onedomain}
\end{figure}

A central yet unresolved issue is the domain structure.
A single-domain stripe phase gives rise to two low-energy satellites.
The experimentally observed four peaks require
the existence of different domains. On the one hand, INS data
on a sample where one kind of domain dominates \cite{mook00}
support an interpretation in terms of one-dimensional (1D) stripes,
recent STM data \cite{mcelr04b} on the other hand do not
find 1D structures but checkerboard-like patterns at the surface.
The clear prediction of our model for the signature of single-domain stripes
and ladders is depicted in Fig.\ \ref{onedomain} for the same
parameters as in Fig.\ \ref{ybco}. Below $\omega_r \! = \! 34$\,meV, only two
satellite peaks are present. Above $\omega_r$, there are two elongated
features with the intensity peaking at their centers. The positions of the
maximum intensity rotate by 90$^\circ$ around $Q_{\rm AF}$ on sweeping
through $\omega_r$, in contrast to the rotation by 45$^\circ$ observed in
multi-domain samples.
The shape and orientation of the features above $\omega_r$ does not depend
strongly on the energy, in contrast to the result for a spin-symmetry
broken phase of the Hubbard model \cite{seibo04}.
To find structures like the ones in Fig.\ \ref{onedomain} would be the
`smoking gun' of quasi-1D spin physics.
Alternatively, for short-range stripe correlations one will observe patterns
as in Fig.\ \ref{ybco}.
At present, there is no theoretical prediction for INS how to distinguish
short-range stripe correlations from the superposition of different
domains of long-range stripes.

In conclusion, a model of coupled spin ladders with established
coupling parameters leads to very good agreement with neutron data.
The predictions made will help to distinguish
stripes from checkerboard patterns or other scenarios.

We gratefully acknowledge the provision of experimental data
for Fig.\ 1 by J.M.\ Tranquada and the
financial support by the DFG in SFB608 and SP1073. One of us
(GSU) thanks the COE at Tohoku University, Sendai, for hospitality and
financial support.

\end{document}